\documentclass{iopart}
\usepackage{epsf}

\jl{6}        
\eqnobysec    

\def\beq{\begin{equation}}
\def\eeq{\end{equation}}

\input prepictex \input pictex \input postpictex
\def\mathput#1{\relax \ifmmode \displaystyle #1\else $\displaystyle #1$\fi}

\begin{document}

\title[Gravitomagnetism and Relative Observer Clock Effects]
{Gravitomagnetism and Relative Observer Clock Effects}

\author{
Donato Bini\dag ${}^\ast$,
Robert T. Jantzen$\S{}^\ast$, 
Bahram Mashhoon\ddag
}

\address{
  \dag\
  Istituto per Applicazioni della Matematica C.N.R.,   I--80131 Napoli, Italy \\
}

\address{
  ${}^\ast$
 I.C.R.A. International Center for Relativistic Astrophysics, University of Rome, I--00185 Rome, Italy
}

\address{
  \S\
  Department of Mathematical Sciences, Villanova University, 
  Villanova, PA 19085, USA
}
\address{
  \ddag\
Department of Physics and Astronomy,
University of Missouri-Columbia, Columbia
MO 65211, USA 
}

\begin{abstract}
The gravitomagnetic clock effect and the Sagnac effect for circularly rotating orbits in stationary axisymmetric spacetimes are studied from a relative observer point of view, clarifying their relationships and the roles played by special observer families. In particular Semer\'ak's recent characterization of extremely accelerated observers in terms of the 
two-clock clock effect is shown to be complemented by a similarly special property of the single-clock clock effect. 
\end{abstract}

\pacs{0420C}

\submitted October  25, 2000  [revised version: November 16, 2000]

\section{Introduction}

Stationary axially symmetric spacetimes, especially the Kerr spacetime as the model for rotating black holes, have proven to be a useful arena for studying various relativistic effects. In particular the consequences of rotation in relativity compared with the simpler picture of Newtonian physics are naturally explored within this class of spacetimes, other interesting cases for this purpose being the G\"odel spacetime in its stationary axially symmetric form and the Minkowski spacetime in uniformly rotating cylindrical coordinates.

Over the past decade a significant effort from various points of view has been invested in developing  a careful analysis of test particle and test gyroscope motion in rotating black hole spacetimes in connection with the definition of inertial forces in general relativity as well as in clarifying the meaning of `nonrotation' in a `rotating' spacetime. The tools of gravitoelectromagnetism \cite{mfg,masgrolic}, i.e.\ spacetime splitting techniques based on families of test observers, can be very helpful in this investigation.
Each such family measures spacetime quantities relative to its own local space and time directions, thus performing a `relative observer analysis.'  
Here we study the so-called `gravitomagnetic clock effect' and the related Sagnac effect using these tools. This is especially interesting for the clock effect since there are actually attempts under consideration to measure it. 

One must distinguish three distinct gravitomagnetic clock effects for a pair of oppositely rotating circular geodesic test particles (oppositely rotating with respect to an intermediate observer):
\begin{enumerate}
\item
the observer-dependent single-clock clock effect: the difference between the periods of the oppositely rotating geodesic test particles as measured by the observer's clock \cite{idcf2};
\item
the observer-dependent two-clock clock effect: the difference between the periods of the oppositely rotating geodesic test particles as measured by their own clocks for one revolution with respect to an observer
\cite{cohmas,bonste,sem}; and 
\item
the observer-independent two-clock clock effect: the difference between the periods of the oppositely rotating geodesic test particles as measured by their own clocks between two crossing events
\cite{mitpul,fermas,tart1,tart2}.
\end{enumerate}
In the first two cases, for a given observer, one compares the periods of one revolution of these orbits (starting from and returning to the same observer world line) measured either by the observer's own clock (single-clock effect) or by the clocks carried along the two orbits (two-clock effect). In the third case, no observer enters the calculation.

Here we draw some connections between these effects and the Sagnac and desynchronization effects and the usual symmetry-adapted coordinates in these spacetimes, while extending some previous work. In particular we extend Semer\'ak's result \cite{sem} 
for the observer-dependent two-clock clock effect for extremely accelerated observers in the Kerr and van Stockum spacetimes to general stationary axially symmetric spacetimes and explain its connection to the single-clock clock effect, where these observers also play a special role. 
Finally the observer-independent two-clock clock effect is illustrated for the G\"odel spacetime and all the clock effects for the equatorial plane of the Kerr-Taub-NUT spacetime.

\section{Metric splitting}

Consider the spacetime metric \beq
ds^2 = ds^2_{(t,\phi)} + g_{rr}dr^2+g_{\theta\theta} d\theta^2 
               \ ,
\eeq
with the metric of the circular orbit cylinder written in the respective coordinate, threading and slicing notation
\begin{eqnarray}
\label{eq:metric}
       ds^2_{(t,\phi)} &=& g_{tt} dt^2 
                    + 2 g_{t\phi} dt d\phi
	              +  g_{\phi\phi} d\phi^2 
                 \nonumber\\
	&=& -M^2 (dt - M_\phi d\phi)^2 + \gamma_{\phi\phi} d\phi^2
                  \nonumber\\
    &=& -N^2 dt^2 + g_{\phi\phi}(d\phi + N^\phi dt)^2
 \ ,
\end{eqnarray}
where in the case of Kerr spacetime, $\{t,r,\theta,\phi \}$ are the usual Boyer-Lindquist coordinates adapted to the family of static observers (or threading observers, with 4-velocity $m=M^{-1}\partial_t$ and world lines along the time coordinate lines) and to the ZAMOs (or slicing observers, with 4-velocity $n=N^{-1}(\partial_t-N^\phi\partial_\phi)$ and world lines orthogonal to the time coordinate hypersurfaces). 
The spacelike unit vectors giving the (increasing $\phi$) angular direction in the local rest space of these observers are 
\beq\label{eq:barmn}
\bar m = \gamma_{\phi\phi}^{-1/2}(\partial_\phi+M_\phi \partial_t) \ ,\
\bar n= g_{\phi\phi}^{-1/2}\partial_\phi\ .
\eeq
The measurement of spacetime quantities in terms of the $1+3$ orthogonal decomposition associated with the threading observers (moving along the time lines---the `threading' of spacetime) is called the threading point of view, while the corresponding measurement process associated with the slicing observers (moving orthogonal to the time hypersurfaces---the `slicing' of spacetime) is called the slicing point of view.

For other spacetimes in this general symmetry
class, like the cylindrically symmetric G\"odel or van Stockum spacetimes \cite{bonste}, for example, a more convenient notation would use cylindrical-like $\rho$-$z$ coordinates instead of spherical-like $r$-$\theta$ coordinates and the Lewis-Papapetrou form of the metric.
In any case, when restricting this metric to the equatorial plane $\theta=\pi/2$ in Kerr, as will be done for most of our explicit work, a simple correspondence $-\partial_\theta \to\partial_z$ and $r\to\rho$ will map one form onto the other.

All of the Kerr-specific discussion below applies to the entire type-D  
many-parameter family of solutions treated by Carter \cite{carter,kramer}. For these spacetimes an additional preferred observer 4-velocity $u_{\rm(car)}$ \cite{znajek} is the direction of the intersection of the $t$-$\phi$ plane in the tangent space with the plane spanned by the two independent principal null directions of the Weyl tensor; in this frame the electric and magnetic parts of the Weyl tensor and of the electromagnetic field tensor if present are proportional 
(the form of the Kerr metric adapted to the Carter observers is just the Boyer-Lindquist difference of squares form of the $t$-$\phi$ 
2-metric given in Misner, Thorne and Wheeler \cite{mtw}, equation~(33.2)).
Furthermore, in this class of spacetimes the timelike Killing vector $\partial_t$ along $m$ is assumed to be the unique such field which has unit norm at spatial infinity in the direction away from the symmetry axis, i.e. corresponding to the distantly nonrotating observers.
For all spacetimes under consideration, the spacelike Killing vector field
$\partial_\phi$ is the unique such field with closed integral curves.
 
The single-clock clock effect discussion 
requires that there exist a pair of oppositely rotating geodesics with respect to the distinct preferred observer $m$, which excludes the case that the Killing direction $m$ itself is geodesic, as occurs in the G\"odel spacetime. However, the two-clock clock effect can still be considered there.

The lapse and shift notation in the two points of view has the following dual correspondence
\beq
\partial_t = Nn +N^\phi \partial_\phi\ , \qquad 
dt= M^{-1}(-m^\flat )+M_\phi d\phi \ ,
\eeq
where $m^\flat$ is the `index-lowered' 1-form corresponding to the vector field $m$.
If $\tau_u$, $u=m,n$, is a proper time parametrization of the threading and slicing world lines, then  
the lapse functions relate  differential increments of coordinate time to proper time along the world line
\begin{equation}\label{eq:lapse}
   d\tau_m = M dt\ ,\qquad
   d\tau_n = N dt\ .
\end{equation}
For constant speed circular orbits, the same relationships apply to finite time intervals.

The shift fields describe the tilting of the local time and space directions of the slicing and threading observers respectively from the time coordinate lines and hypersurfaces. In the slicing point of view the local time direction of the slicing observers is the direction for which the 1-form $\bar n^\flat\propto d\phi +N^\phi dt$ vanishes 
(equivalent to orthogonality to the angular direction $\bar n$)
\beq
    d\phi = -N^\phi dt \to d\phi/dt = -N^\phi 
    =\zeta_{\rm(sl)}\ ,
\eeq
which gives the rate of change of angle with respect to time (angular velocity) of the slicing observer world line following the local time direction.

In the threading point of view, the local spatial angular direction (in the local rest space of the threading observer) is the direction for which the 1-form $m^\flat\propto   dt - M_\phi d\phi$ vanishes
(equivalent to orthogonality to the direction $m$)
\beq\label{eq:dtM}
dt = M_\phi d\phi \to dt/d\phi = M_\phi
   =\bar\zeta_{\rm(th)}{}^{-1}\ ,
\eeq
which gives the rate of change with respect to angle of the time coordinate along a circular curve which is spatial with respect to the threading observers (an inverse angular velocity). The coordinate time difference which accumulates along such a curve after one complete revolution with respect to the threading observers, which may be converted into a proper time with respect to those observers by multiplication by the lapse function, is called the synchronization gap \cite{baz}
\beq
  \Delta t_{\rm(SG)} = 2\pi M_\phi\ ,\qquad 
  \Delta \tau_{\rm(SG)} = 2\pi M M_\phi\ . 
\eeq

Circular orbits in the Kerr spacetime equatorial plane will be used to illustrate the various ideas introduced for the entire symmetry class of spacetimes under consideration.
The metric coefficients, lapse and shift factors, and angular velocities of the local time and angular directions for this case are given explicitly in Table 1. 

\typeout{********* TABLE OF KERR FUNCTIONS}

\typeout{*** Table 1. (thdsli)}
\begin{table}\footnotesize
\caption
{The lapse 
and the nonzero observer-adapted components of the shift
and spatial metric in both the threading and slicing points of view
are given for the Kerr black hole spacetimes on
the equatorial plane, with the abbreviation
$ \Delta= r^2 -2{\cal M} r + a^2$. The assumption $a>0$ corresponds
to the hole
rotating in the counterclockwise (forward) angular $\phi$-direction. Angular velocities determining the local time direction and local forward angular direction are also given for these observers and the Carter observers. 
}
\vbox{%
 \def\hline{}
 \typeout{*** eqnarray struts inserted}
 \def\Strut{\relax\hbox{\vrule width0pt height 10.5pt depth 5.5pt}}
\begin{eqnarray*}
\begin{array}{ll} 
 \hline \Strut
N= (-  g^{tt})^{-1/2}       
                           & \sqrt{r\Delta/(r^3+a^2r+2a^2{\cal M})}
                      \\ \hline \Strut
N^\phi    
                         & -2a{\cal M}/(r^3 + a^2 r +2a^2{\cal M})
                        \\ \hline \Strut
g_{\phi\phi}  
                                & (r^3 + a^2r + 2a^2{\cal M})/r
                        \\ \hline \Strut
M= (-  g_{tt})^{1/2}        
                                & \sqrt{(r-2{\cal M})/r} 
                         \\ \hline \Strut
M_\phi = -   g_{t\phi}/  g_{tt} 

                                & -2a{\cal M}/(r-2{\cal M}) 
                          \\ \hline \Strut
\gamma_{\phi\phi} =   g_{\phi\phi}
                  -(  g_{t\phi})^2
                   /  g_{tt}
                                & r\Delta/(r-2{\cal M})  
                          \\ \hline \Strut
g_{rr}  
                                & r^2/\Delta 
                         \\ \hline \Strut
g_{\theta\theta}  
                                & r^2
                         \\ \hline \Strut
\zeta_{\rm(sl)}\ ,\ \bar\zeta_{\rm(sl)} 
                                & -N^\phi \ ,\ 0
                         \\ \hline \Strut
\zeta_{\rm(th)}\ ,\ \bar\zeta_{\rm(th)} 
                                & 0 \ ,\ M_\phi^{-1}                                                   \\ \hline \Strut
\zeta_{\rm(car)}\ ,\ \bar\zeta_{\rm(car)} 
                                & a/(r^2+a^2) \ ,\ 1/a                        \\ \hline
\end{array}
\end{eqnarray*}
}
\label{tab:thdsli}
\end{table}

\section{Circular orbit geometry}

Consider a generic observer with 4-velocity $U$ following a circular orbit world line, moving along the $\phi$ direction with constant angular velocity $\zeta=d\phi/dt =\dot\phi$. Its world lines may be taken as a new threading associated with the new comoving coordinates defined by
\beq
     \tilde t = t\ ,\ \tilde\phi = \phi-\zeta t\ .
\eeq
The orthogonal decomposition of the metric adapted to this observer corresponds to the new threading decomposition
\beq
\label{eq:Umetric}
       ds^2_{(t,\phi)} 
     = -\tilde M{}^2 (d\tilde t - \tilde M_{\tilde\phi} d\tilde\phi)^2
            + \tilde\gamma_{\tilde\phi\tilde\phi} d\tilde\phi{}^2
 \ ,
\eeq
where the new threading lapse and shift fields are
\beq
  \tilde M = \Gamma^{-1} 
   \ ,\ 
  \tilde M_{\tilde\phi} = (g_{\phi\phi} \zeta+ g_{t\phi}) \Gamma^2
   = (\bar\zeta-\zeta)^{-1}
\ .
\eeq

The new observer 4-velocity $U$ and its orthogonal partner $\bar U$ in both sets of coordinates are
\begin{equation}
\fl\qquad
\label{eq:zeta}
   U
    = \Gamma ( \partial_t + \zeta \partial_\phi ) 
    =\tilde M{}^{-1} \partial_{\tilde t}
      \ ,\quad
   \bar U 
    = \bar\Gamma ( \partial_t + \bar\zeta \partial_\phi )
    = \tilde\gamma{}_{\tilde\phi\tilde\phi}^{-1/2} 
       (\partial_{\tilde\phi} + \tilde M_{\tilde\phi} \partial_{\tilde t})
      \ .
\end{equation}
The quantity $\bar\zeta{}^{-1}$ is the rate of change of coordinate time per angle along the spatial angular direction of this new observer
(along which $dt-\bar\zeta{}^{-1} d\phi=0$), associated with the desynchronization effect.
The `coordinate gamma factor' 
$\Gamma = dt/d\tau_U  >0$ is defined by
\begin{eqnarray}\label{eq:gammasqd}
  \Gamma^{-2} 
    &=& 
     -[            g_{tt} 
         + 2 \zeta g_{t\phi} 
         + \zeta^2 g_{\phi\phi}]
    =         - g_{\phi\phi}
            (\zeta - \zeta_-)(\zeta - \zeta_+)
              \nonumber\\
    &=&  M^2(1- M_\phi \zeta)^2 - \gamma_{\phi\phi} \zeta^2 
    =  N^2 - g_{\phi\phi}(\zeta + N^\phi)^2 \ .
\end{eqnarray}
The reciprocal $\tilde M=\Gamma^{-1}$ is the lapse function relating increments of coordinate time to a proper time parameter $\tau_U$ along the new threading world line 
\begin{equation}\label{eq:lapseparticle}
   d\tau_U = \Gamma^{-1} dt\ ,
\end{equation}
and for constant speed circular orbits, the same relationship applies to finite time intervals.

The
timelike condition for the unit 4-velocity $U$ requires $\Gamma^{-2}
> 0$, constraining $\zeta$ to belong to the interval
$(\zeta_-,\zeta_+)$ between the roots of the quadratic equation
$\Gamma^{-2}=0$ in $\zeta$ corresponding to null directions, namely 
\begin{equation}\label{eq:zetapm}
\fl
 \zeta_\pm 
   = [-g_{t\phi} 
           \pm (g_{t\phi}{}^2 - g_{\phi\phi}g_{tt})^{1/2}]
                /g_{\phi\phi} 
   = [M_\phi \pm M^{-1}\gamma_{\phi\phi}^{1/2}]^{-1} 
   = -N^\phi \pm N g_{\phi\phi}^{-1/2} \ .
\end{equation}
A zero-rest-mass test particle (photon) has instead a 4-momentum vector of the same form 
\begin{equation}
 P_\pm 
    = \Gamma_{\rm(null)}
          ( \partial_t + \zeta_\pm \partial_\phi ) \ .
\end{equation}
but with an arbitrary normalization factor
$\Gamma_{\rm(null)}$.
Circular orbits for which $\zeta > 0$ or $\zeta<0$ will be referred
to, respectively, as corotating or counterrotating (with respect to the
`static observers' of the original coordinate grid). 

Figure 1 shows the $t$-$\phi$ plane in the tangent space 
at a typical radius (where both circular geodesics are timelike) in the equatorial plane in the region of Kerr spacetime
superimposed on the $t$-$\phi$ cylinder, with its tilted vertical axis along $t$, tilted left with respect to the axis along $n$ shown perpendicular to the $\phi$-direction. The null vectors $P_\pm = n\pm \bar n$ are shown with the normalization $\Gamma_{\rm(null)} = N^{-1}$, where $\bar n$ 
is a unit vector along the $\phi$-coordinate line. The other two angular directions $\bar m$ and $\bar u_{\rm(car)}$ are also shown at the end of one revolution of a curve which is spatial with respect to the associated observer. 
The coordinate time difference $\Delta t$ 
after one such revolution $\phi: 0\to 2\pi$ along 
the curve with direction $\bar m$ 
is called the synchronization gap for 
the static observer $m$, easily converted into a proper time difference measured by this observer.

The physical components of the velocities measured by the threading
and slicing observers for motion along circular orbits 
are related to the coordinate
angular velocity by linear or fractional linear transformations
\begin{eqnarray}\label{eq:nuUmn}
\fl\qquad
  \nu(U,m)^{\hat\phi} 
     = \gamma_{\phi\phi}{}^{1/2} \zeta / [M(1-M_\phi \zeta)] \ ,\quad 
  \nu(U,n)^{\hat\phi} 
     = g_{\phi\phi}{}^{1/2} (\zeta + N^{\phi})/N
\end{eqnarray}
and
\begin{eqnarray}\label{eq:zetanuUmn}
\fl\qquad
   \zeta 
  = M \nu(U,m)^{\hat\phi}/ 
      [\gamma_{\phi\phi}{}^{1/2} + M M_\phi \nu(U,m)^{\hat\phi}]
  = -N^\phi + N g_{\phi\phi}{}^{-1/2}  \nu(U,n)^{\hat\phi} \ .
\end{eqnarray}
Here the hatted index
notation $\hat\phi$ refers to the component along the 
unit vector $\bar m$ or $\bar n$ of (\ref{eq:barmn}) giving the forward (increasing $\phi$) angular direction in the local rest space of each point of view.
Note that when the shift is nonzero, test particle motions with
angular velocities of equal magnitude but opposite sign lead to
physical velocities which do not have the same magnitude and vice
versa. When $\nu(U,u)^{\hat \phi} = \pm 1$, the latter equation
reduces to Eq.~(\ref{eq:zetapm}).  
The dashed line in Figure 1 between the tips of $P_\pm$ in the tangent space contains the slicing relative velocities associated with the various observer world lines as the connecting vectors between the tip of $n$ and the points of intersection of those world lines (actually their 4-velocity vectors) with the dashed line.

The `coordinate' gamma factor is easily expressed in terms of the
usual Lorentz gamma factor associated with these relative velocities 
\begin{eqnarray}\label{eq:gammagamma}
     \Gamma &=& \gamma(U,m)/ [ M(1-M_\phi \zeta)] \equiv \Gamma(U,m) 
                  \nonumber\\
            &=& \gamma(U,n)/ N \equiv \Gamma(U,n) \ .
\end{eqnarray}
These formulas may be used to express the angular momentum (per unit
mass) 
\begin{equation}\label{eq:angmom}
  p_\phi = U_\phi 
  = \Gamma\nu(U,n)_\phi 
  =  \Gamma g_{\phi\phi} ( \zeta-\zeta_{\rm(sl)} )
\end{equation}
of $U$ defined by the rotational Killing vector $\partial_\phi$
and its Killing energy (per unit mass) ${\cal E} = -U_t =
M^{-1}\gamma(U,m)$ defined by the Killing vector $\partial_t$,
both conserved for geodesic motion. These are related to the
coordinate gamma factor by the identity $-1=U_\alpha U^\alpha = \Gamma
(-{\cal E} +\zeta p_\phi)$ in the timelike case and $0 =P_\alpha
P^\alpha = \Gamma_{\rm(null)} (-{\cal E} +\zeta p_\phi)$ in the null
case, where $({\cal E},p_\phi)=(-P_t,P_\phi)$. 
In the timelike case the ratio
\begin{equation}\label{eq:barU}
  \bar\zeta = \frac{{\cal E}}{p_\phi}
            = -\frac{g_{tt}+\zeta g_{\phi t}}
                    {g_{t\phi}+\zeta g_{\phi\phi}}
\end{equation}
defines the angular velocity of the circular orbit along $\bar U$ orthogonal to $U$. For the null case, one has the limiting case $\bar\zeta_\pm=\zeta_\pm$ given by this same formula.

\typeout{!!!!!!!!!!!!!!
Figure 1 cannot be shrunk without making it unreadable, and must be on a full page by itself, with its caption on the adjoining page (depending on even odd page placement) at the top or bottom according to your style.
!!!!!!!!!!!!!!}


\begin{figure}[b]
\caption{
This spacetime diagram shows the Sagnac and clock effects in the equatorial plane of the Kerr spacetime at a typical radius where both circular geodesics (labeled by $U_+$ and $U_-$) are still timelike.
The forward and backward circular photon orbits
(labeled by $P_+$ and $P_-$) 
are at a $45^\circ$ angle with respect to the horizontal direction. 
The front half ($-\pi/2\leq \phi\leq \pi/2$) of the $t$-$\phi$ cylinder coordinate grid is shown flattened out and tilted back with respect to the vertical direction along $n$ 
(and the dashed vertical line on the right edge of the plot)
shown perpendicular to the horizontal $\phi$ coordinate axis (constant $t$),
itself aligned with the unit vector $\bar n$ along the slicing spatial angular direction.  The unit vector $\bar m$ is along the threading local rest space angular direction. Extending this direction counterclockwise around one loop of the cylinder leads to the change in the coordinate time equal to the synchronization gap for $m$. Doing the same for the spatial direction $\bar u_{\rm(car)}$ for the Carter observer leads to the single-clock clock effect $\Delta t_{\rm(geo)}$ for $m$. The world lines of the various observers are labeled by their 4-velocities, and their successive crossing points with the pair of circular geodesics and with the pair of photon orbits characterize the clock and Sagnac effects respectively for the given observer.
}
\end{figure}

\begin{figure}[p]
\typeout{pictex  figure 1}
$$ \vbox{
\beginpicture
  \setcoordinatesystem units <0.8cm,0.8cm> point at 0 0 

    \putrule from  -4.575 23   to 3.425 23 
    \putrule from  -3.975 -1 to 4.025 -1   
    \putrule from  -4 0   to 5 0           
    \putrule from  0 0   to 0 23           
    \putrule from  0 0   to 0 -1           

\setdashes

    \putrule from 3.425 23 to 4 23 
    \putrule from 4 19.32  to 4 23      
    \putrule from 4 0  to 4 18.33       


\setlinear

\setsolid                    

  \plot 0 0 -0.625 24 /          
  \plot 0 0  -1 0.0075   /       

  \plot -4 0 -4.575 23 /      
  \plot  4 0  3.425 23 /      

  \plot -4 0 -3.975 -1 /      
  \plot  4 0 4.025 -1  /      

  \plot 0 0 4 -0.1 /          
  \plot -4 -0.3 0 -0.4  /     
  \plot 0 -0.4 2.0625 -0.4525 / 

\plot 0 0 4.0  .25 /              
\plot -4 .75 0 1.0 /              


\setlinear
\setsolid                    
  \plot 0 0 2 2 /               
  \plot 0 0   -2 2 /            

\setsolid


\setdashes 
\setquadratic


\plot
0 2 -0.4 2.03961 -0.8 2.15407 -1.2 2.33238 -1.6 2.56125 
-2 2.82 -2.4 3.1241 -2.8 3.44093 -3.2 3.77359 -3.4 3.94462
-3.8 4.29418  /

\plot
0 2 0.4 2.03961 0.8 2.15407 1.2 2.33238 1.6 2.56125 
2 2.82 2.4 3.1241 2.8 3.44093 3.2 3.77359 3.4 3.94462
3.8 4.29418 /

\setlinear
\setdashes

  \plot  2 2   3.9 3.9  /       
  \plot  -2 2   -4.1 4.1 /      

  \plot -4.25 11.75 3.5 19.5 /  
  \plot 3.7 12.3 -4.5125 20.5 / 

\setsolid

\plot 0 0 -3.88 4.37  /              
\plot 0 0 -4.1 4.5817 /              
\plot 3.7 13.6 -4.575 22.9 /         

\plot 0 0  2.67 3.33  /              
\plot 0 0  3.9 4.9    /              
\plot -4.3 14.6 2.42 23 /            

\plot 0 0 -1.115 23      /           
\plot 0 0 -3.62 23      /            
\plot 0 0 -1.21 7.70 /               

\plot 0 0 3.6 15.5 /                 
\plot 0 0 -.174 -1.0 /               

\setdashes <2pt>


\plot -3.88 4.37 -1.21 7.70 /        
\plot 2.67 3.33  -1.21 7.70 /        
\plot -2 2 2 2 /                     



\put {\mathput{\bullet}}                at  -2.62 16.68
\put {\mathput{\bullet}}                at  0 19.95
\put {\mathput{\bullet}}                at  -0.5 19.32
\put {\mathput{\bullet}}                at  -0.9 18.83
\put {\mathput{\bullet}}                at  -0.45 18.30
\put {\mathput{\bullet}}                at  0 17.8
\put {\mathput{\bullet}}                at  -3.375 21.55


\put {\mathput{\bullet}}                at  -3 19
\put {\mathput{\bullet}}                at  -0.83 16.87
\put {\mathput{\bullet}}                at  -0.4 16.4
\put {\mathput{\bullet}}                at  0 16
\put {\mathput{\bullet}}                at  -2.18 13.82
\put {\mathput{\bullet}}                at  -0.737 15.22
\put {\mathput{\bullet}}                at  -0.3725 15.63

\put {\mathput{\bullet}}                at 0 1.0



\put {\mathput{\bullet}}                at  0 0
\put {\mathput{\bullet}}                at  0 -0.4


\plot -3.375 21.55 3.45 21.55 / 
\plot 0 19.95 3.4 19.95 /      
\plot -0.5 19.32 3.515 19.32 / 
\plot -0.9 18.83 3.53 18.83 /  
\plot -0.45 18.3 3.55 18.3 /   
\plot  0 17.8 3.57 17.8 /      
\plot -2.6 16.7 3.6 16.7 /     

\setdashes <4pt>


\plot -4 -0.4 0 -0.4 /              


\plot -4.47 19 -3 19 /              
\plot -4.41 16.87 -0.83 16.87 /     
\plot -4.4 16.4 -0.4 16.4 /         
\plot -4.385 16 0 16 /              
\plot -4.3725 15.63 -0.3725 15.63 / 
\plot -4.36 15.22 -0.737 15.22 /     
\plot -4.32 13.82 -2.18 13.82 /     

\setsolid

\put {\mathput{n}}                  [b]   at  0 23.25
\put {\mathput{m}}                  [lb]   at  -0.4 24
\put {\mathput{U_{\rm (gmp)}}}      [rb]   at  -0.7 23.25
\put {\mathput{U_{\rm (ext)}}}      [b]   at  -3.5 23.25
\put {\mathput{U_-}}                [b]   at  -4.575 23.25
\put {\mathput{U_+}}                [b]   at  2.5 23.25
\put {\mathput{U_-}}                [lb]   at  -3 4
\put {\mathput{U_+}}                [rt]   at  2 3.5
\put {\mathput{U_+ + U_-}}          [rb]   at  -1.7 7.5
\put {\mathput{P_+}}                [lt]   at  -3.5 12
\put {\mathput{P_-}}                [rt]   at   2.0 13.7 
\put {\mathput{n}}                  [rt]   at   -0.6 1.75  
\put {\mathput{P_+}}                [lt]   at  2.25 1.75 
\put {\mathput{P_-}}                [rt]   at  -2.25 1.75 
\put {\mathput{\bar u_{\rm (car)}}} [rt]   at   3 1.0 
\put {\mathput{\bar m}}             [l]   at  2.5 -0.5
\put {\mathput{\bar n}}             [rt]   at   1.5 0.4
\put {\mathput{t}}                  [rb]   at  -1 24
\put {\mathput{\phi}}               [b]   at  4.5 -0.75
\put {\mathput{-\frac{\pi}{2}}}     [rb]   at  -4.15 0.3
\put {\mathput{\frac{\pi}{2}}}      [lb]   at   4.15 0.3
\put {\mathput{\Delta t_{\rm (SG)}}}   [r]   at -4.15 -0.2
\put {\mathput{u_{\rm (car)}  }}         [l] at  1.2 10


\betweenarrows 
  {\mathput{\Delta t_{\rm (geo)}}} [l] from 3.65 17.03 to 3.65 20.63
\betweenarrows 
  {\mathput{\Delta t_{\rm (null)}}} [r] from -4.5 14.5 to -4.5 17.5

\arrow <.5cm> [.1,.4]    from  4.5 0 to  5  0               
\arrow <.5cm> [.1,.4]    from  -0.6250 24 to -0.6275 24.5   

\arrow <.5cm> [.1,.4]    from  1.5 0 to  2  0              
\arrow <.5cm> [.1,.4]    from  1.5 -0.45 to 2.0625 -0.4525 
\arrow <.5cm> [.1,.4]    from  0 1.5 to  0 2               
\arrow <.5cm> [.1,.4]    from  0.0 1.0 to 2.1 1.13         

\arrow <.5cm> [.1,.4]    from   0 0  to    2.67 3.33           
\arrow <.5cm> [.1,.4]    from   0 0  to -3.88 4.37             
\arrow <.5cm> [.1,.4]    from   1.65 1.65 to    2 2            
\arrow <.5cm> [.1,.4]    from  -1.65 1.65 to   -2 2            
\arrow <.5cm> [.1,.4]    from  0 0 to -1.21 7.70               

\setsolid
\setlinear 
\linethickness=0.8pt
\plot -0.45 18.30  -0.48 19.32 /
\plot -0.38 15.63  -0.41 16.4 /

\endpicture}$$

\end{figure}

Note from Eq.~(\ref{eq:zetapm}) that the coordinate angular velocity of the slicing observers is
just the average of the two limiting angular velocities 
\begin{equation}\label{eq:sagnacobservers}
     \zeta_{\rm(sl)}  
      = \zeta_{\rm(nmp)} 
      = (\zeta_- + \zeta_+)/2 
      = - N^\phi\ ,
\end{equation}
which is just equation (33.16) of Misner, Thorne, and Wheeler
\cite{mtw} in the specific context of the Kerr spacetime. Their
exercise (33.3) following the discussion of Bardeen \cite{bar} applies
to the general (orthogonally transitive) stationary axially symmetric
case \cite{greschvis}, so that one may interpret the slicing observers
as the locally nonrotating observers with respect to the Sagnac
effect \cite{pos,ash,hen,andbilste,ste1,ste2}. They experience no Sagnac effect for the oppositely directed
accelerated photons constrained by mirrors or fiber optical cable to
remain on a given circular orbit, meaning that the alternating meeting
points of these photons (`null meeting points', occurring every other crossing of their world lines) lie on the same observer world line. They are
also called the `zero angular momentum observers' 
(ZAMO's), since they
are orthogonal to the angular Killing vector and therefore have
vanishing angular momentum. 

Similarly the spatial $\phi$ direction with respect to the original threading observers is defined by the angular velocity $\bar\zeta_{\rm(th)} = M_\phi{}^{-1}$. By direct calculation from (\ref{eq:zetapm}) one finds the following result 
\begin{equation}\label{eq:threadingshift}
\bar\zeta_{\rm(th)}{}^{-1}
     = (\zeta_-{}^{-1} + \zeta_+{}^{-1})/2 = M_\phi 
\ .
\end{equation}
This establishes a complementary relation between the null angular velocity reciprocals and the threading shift 1-form coordinate component in direct analogy to the slicing relation between the angular velocities themselves and the shift vector coordinate component (\ref{eq:sagnacobservers}). 

\begin{figure}
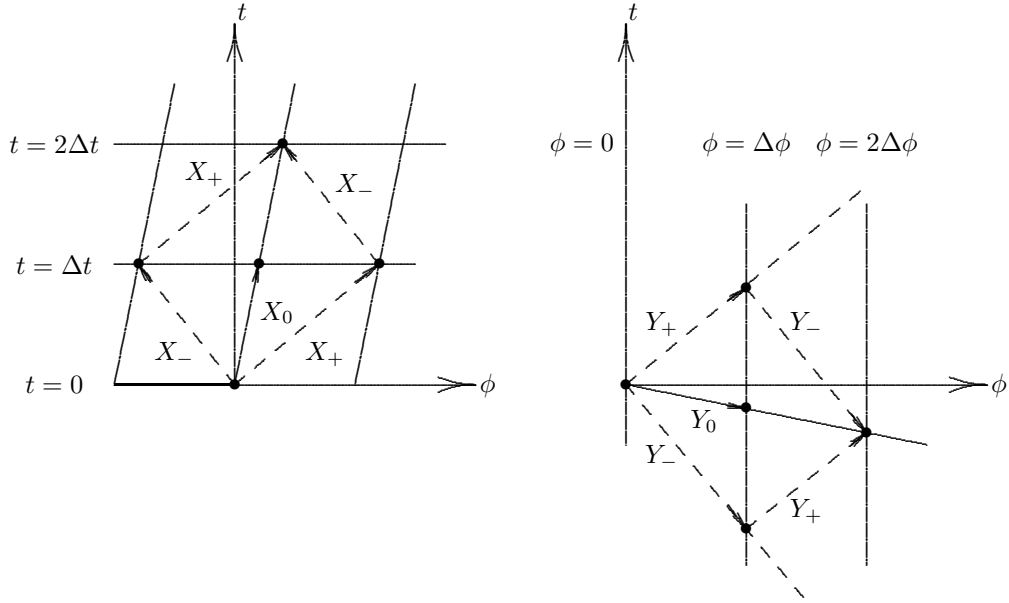

\typeout{pictex  figure 2}
\vspace{2cm}
$$ \vbox{
\beginpicture
  \setcoordinatesystem units <0.8cm,0.8cm> point at 0 0 


\putrule from  -2 0   to 0 0                
\arrow <.5cm> [.1,.4]    from  0 0 to 4 0   
\arrow <.5cm> [.1,.4]    from  0 0 to 0 6   

\arrow <.25cm> [.15,.3]    from   2 1.7 to 2.4 2 
\arrow <.25cm> [.15,.3]    from   -1.4 1.7 to -1.6 2 
\arrow <.25cm> [.15,.3]    from  0.5 3.75 to  0.8 4
\arrow <.25cm> [.15,.3]   from  1 3.75 to 0.8 4
\arrow <.25cm> [.15,.3]   from  0.35 1.75 to 0.4 2

\setdashes
\plot 0 0    2.4 2  /               
\plot 0 0   -1.6 2  /               
\plot 2.4 2  0.8 4  /               
\plot  -1.6 2 0.8 4 /               
\setsolid

\plot -2 0    -1 5 /              
\plot  0 0    1 5 /               
\plot  2 0    3 5 /               
\plot -2 2    3 2 /               
\plot  -2 4    3.5 4 /            

\put {\mathput{t}}                    [r]  at  0.2 6.2
\put {\mathput{\phi}}                 [b]  at  4.2 -0.2
\put {\mathput{\bullet}}                at  0 0
\put {\mathput{\bullet}}                at  2.4 2
\put {\mathput{\bullet}}                at  0.4 2 
\put {\mathput{\bullet}}                at  -1.6 2
\put {\mathput{\bullet}}                at   0.8 4
\put {\mathput{t=0}}                      at  -3 0
\put {\mathput{t=\Delta t}}             at  -3 2
\put {\mathput{t=2\Delta t}}            at  -3 4
\put {\mathput{X_+}}                    at  1.5 0.5
\put {\mathput{X_+}}                    at  -0.5 3.5
\put {\mathput{X_-}}                    at  -1 0.5
\put {\mathput{X_-}}                    at  2 3.3
\put {\mathput{X_0}}                    at  0.7 1.2


\setcoordinatesystem units <0.8cm,0.8cm> point at -6.5 0 
        
\arrow <.5cm> [.1,.4]    from  0 0 to 0 6   
\arrow <.5cm> [.1,.4]    from  0 0 to 6 0   

\arrow <.25cm> [.15,.3]    from   1.7 1.4 to 2 1.6  
\arrow <.25cm> [.15,.3]    from   1.7 -2 to 2 -2.4
\arrow <.25cm> [.15,.3]    from  3.75 -0.5  to   4 -0.8
\arrow <.25cm> [.15,.3]    from  3.75  -1 to 4 -0.8
\arrow <.25cm> [.15,.3]    from  1.7  -0.35 to 2 -0.4

\setdashes 
\plot  0 0    4 3.3  /                 
\plot  0 0    3 -3.6  /                
\plot  2 1.6    4 -0.8  /              
\plot  2 -2.4  4 -0.8  /               
\setsolid
\plot  0 0    0 -1  /               
\plot  2 -3   2 3  /                
\plot  4 -3   4 3  /                
\plot 0 0   5 -1 /                  

\put {\mathput{t}}                 [r]  at  0.2 6.2
\put {\mathput{\phi}}              [b]  at  6.2 -0.2
\put {\mathput{\bullet}}                at  0 0
\put {\mathput{\bullet}}                at  2 1.6
\put {\mathput{\bullet}}                at  2 -0.4
\put {\mathput{\bullet}}                at  2 -2.4
\put {\mathput{\bullet}}                at  4 -0.8 
\put {\mathput{\phi=0}}               [r] at  -0.2 4
\put {\mathput{\phi=\Delta \phi}}       at  2 4
\put {\mathput{\phi=2\Delta \phi}}      at  4 4
\put {\mathput{Y_+}}                    at   0.6 1
\put {\mathput{Y_+}}                    at   3 -2.1
\put {\mathput{Y_-}}                    at  0.6 -1.2
\put {\mathput{Y_-}}                    at  3 1
\put {\mathput{Y_0}}                    at  1.3 -0.6
\endpicture}$$
\vspace{1cm}

\caption{
The null synchronization process associated with the interpretation of the shift fields in the slicing and threading points of view. Identifying nearby events with points in the tangent space, corotating ($+$) and 
counterrotating ($-$) photon paths are marked by dashed lines. In the left diagram, the slicing observers are locally synchronized by the $\phi$ coordinate lines, while in the right diagram the threading observers are locally synchronized orthogonally to the $t$-coordinate lines, both within the $t$-$\phi$ cylinders, as explained in the text. The null vectors and their averages are given by
\begin{eqnarray*}
X_\pm &=& \Delta t (\partial_t + \zeta_\pm \partial_\phi) \ ,\ 
X_0 =(X_+ +X_-)/2\ ,\\
Y_\pm &=& \Delta \phi (\partial_\phi + \bar \zeta{}^{-1}_\pm \partial_t)\ ,\
Y_0 =(Y_+ +Y_-)/2\ .
\end{eqnarray*}
}
\end{figure}

Both of these relationships are associated with the null path synchronization of nearby observers illustrated in Figure 2, where
nearby slicing observer world lines are identified with paths in the tangent space.
In the left diagram, the slicing observers are defined so that constant time coordinate lines are synchronized.
The null paths (photon world lines) starting at the origin from the middle observer at $t=0$ are reflected from either of the nearby observers at time $t=\Delta t$ and return to the middle observer at time $t=2\Delta t$. This synchronizes the reflection events with the location of the middle observer at time $t=\Delta t$, at the event identified with the tip of the average vector $X_0$ in the tangent space, and leads to the angular velocity of the middle slicing observer being equal to the average of the photon angular velocities, yielding the formula (\ref{eq:sagnacobservers}) for the sign-reversed angular shift vector component. In the right diagram, the threading observers instead follow  lines of constant $\phi$ in the tangent space, and synchronize themselves along the orthogonal direction. The middle observer at $\phi=\Delta\phi$ synchronizes the origin on the left observer at $\phi=0$ with the event $Y_0$ on his world line corresponding to half the photon round trip to and from the left observer. This leads to the inverse angular velocity $\bar\zeta_{\rm(th)}{}^{-1}$ of the spatial synchronization direction being the average of the corresponding inverse angular velocities $\bar\zeta_\pm^{-1} =\zeta_\pm^{-1}$ of the photons, yielding the angular shift 1-form component formula (\ref{eq:threadingshift}).

The threading discussion also applies to the new threading of the generic observer. This diagram shows how the threading shift characterizes the desynchronization effect, which is nothing more than the local tangent space version of the Sagnac effect through the local photon synchronization process. The coordinate time difference  $\Delta t$ between the 
$\phi$-coordinate line and the right observer at the event $2Y_0$, extrapolated along the two oppositely directed photon circular orbits, becomes the Sagnac effect, just twice the extrapolation of the synchronization gap between the $\phi$ coordinate line and the event $Y_0$. This result, which in turn is closely connected with the various clock effects,
will be shown explicitly below.

\section{Other special observer families}

In the same way that the slicing observers are characterized as the `null meeting point' observers, one can introduce the `geodesic meeting point' observers \cite{mitpul,fermas,mit} that see no time delay between the arrival of oppositely rotating circular geodesics, i.e. the same period for the two orbits as seen by that observer. Their angular velocity, like in the null case, is just the average of the oppositely-signed angular velocities $\dot\phi_- <0< \dot\phi_+$ of these two circular orbits
\begin{equation}
  \zeta_{\rm(gmp)} = (\dot\phi_+ + \dot\phi_-)/2 
       = (|\dot\phi_+| - |\dot\phi_-|)/2 \ ,
\end{equation}
as a short calculation shows 
($\dot\phi_+ t = 2\pi +\Delta\phi$,
$-\dot\phi_- t = 2\pi -\Delta\phi$, subtract and solve for $\Delta\phi$ to
find coefficient of $t$ as the desired angular velocity).
This meeting point condition is valid for any pair of oppositely-signed angular velocities $(\zeta_1,\zeta_2)$, $\zeta_1 <0< \zeta_2$, so that one can introduce
\begin{equation}
  \zeta_{\rm(mp)} = (\zeta_1 + \zeta_2)/2 
       = (\zeta_2 - |\zeta_1|)/2 \ .
\end{equation}
If $U_1$, $U_2$, $U_{\rm(mp)}$
are the corresponding 4-velocities, then the meeting point observers see $U_1$ and $U_2$ moving with angular velocities of equal magnitude but opposite sign
\beq
\zeta_1 - \zeta_{\rm(mp)}  = - [\zeta_2 - \zeta_{\rm(mp)}]\ .
\eeq
Because of the linear relation (\ref{eq:nuUmn}) between $\zeta$ and $\nu(U,n)$, the corresponding slicing relative velocities are also in the same averaging relation
\begin{equation}
  \nu(U_{\rm(mp)},n) = (\nu(U_1,n) +\nu(U_2,n))/2\ .
\end{equation} 
On the other hand differences of angular velocities are proportional to differences of slicing relative velocities for the same reason
\begin{equation}\label{eq:nuzetadiffs}
  \nu(U_1,n)-\nu(U_2,n)= N^{-1}g_{\phi\phi}^{1/2}(\zeta_1 - \zeta_2)\ .
\end{equation} 

\typeout{*** Table 2.}

\begin{table}\footnotesize
\def\hline{} 
\caption
{Causal restrictions 
and geodesic and null conditions for circular orbits
in the Kerr spacetime on the equatorial plane.
} 
 \typeout{*** eqnarray struts inserted}
 \def\Strut{\relax\hbox{\vrule width0pt height 10.5pt depth 5.5pt}}
 \def\TopStrut{\relax\hbox{\vrule width0pt height 10.5pt depth 0pt}}
 \def\BotStrut{\relax\hbox{\vrule width0pt height 0pt depth 5.5pt}}
\begin{eqnarray*}
\begin{array}{ll} \hline 
\hline \TopStrut
 \vtop{\hbox{threading region}\hbox{of validity}}\qquad   
  &r> r_{\rm(erg)}=2{\cal M}\\
  &\\ 
\hline \TopStrut
 \vtop{\hbox{slicing region}\hbox{of validity}}\qquad   
  & r> r_{\rm(h)}={\cal M} +\sqrt{{\cal M}^2 - a^2} \\ \BotStrut
 &\\ 
\hline \Strut
 \hbox{timelike geodesics} 
    &\dot{\phi}_\pm = \frac{\pm\sqrt{{\cal M}/r^3}}
                                  {1\pm a\sqrt{{\cal M}/r^3}}
       \\ \Strut
    &\nu(U_\pm,m)^{\hat{\phi}} 
      = \frac{\sqrt{\Delta}}{a\pm (r-2{\cal M})\sqrt{r/{\cal M}}}
  	\\ \Strut
  &\nu(U_\pm,n)^{\hat{\phi}} = \frac{a^2 \mp 2a\sqrt{{\cal M} r} 
         + r^2}{\sqrt{\Delta}(a\pm r\sqrt{r/{\cal M}})}  \\
\hline \Strut
 \hbox{null orbits} 
    &\zeta_\pm = \frac{2a{\cal M} \pm r\sqrt{\Delta}}
                               {r^3 + a^2 r + 2a^2 {\cal M}} \\
\hline
\end{array}
\end{eqnarray*}	
\label{tab:summaryK}
\end{table}

Another family of preferred observers associated with the clock effects is the one for which the magnitude of the acceleration is extremal with respect to the angular velocity parameter of the circular orbit family \cite{def,sem0,page,idcf2}. 
In the equatorial plane in the Kerr spacetime, one has the inequalities
\beq
  \zeta_{\rm (ext)} < \zeta_{\rm (gmp)} < 0 < \zeta_{\rm (nmp)}\ .
\eeq
The first inequality is discussed in Eqs.~(4.24) and (4.25) of \cite{idcf2}, while an explicit formula for the extremely accelerated observer angular velocity has been given by Semer\'ak \cite{sem}
\beq
  \zeta_{\rm (ext)}
   = \frac{r^2(r-3{\cal M})-2{\cal M}a^2-\sqrt{r^4(r-3{\cal M})^2-4{\cal M}a^2r^3}}{-2{\cal M}a(3 r^2 + a^2)}
\ ,
\eeq
first stated for the variable $y=\zeta/(1-a\zeta)$ by de Felice \cite{def,def0}.
Table 2 leads to the explicit simple formula
\beq
   \zeta_{\rm (gmp)} = - a{\cal M}/r^3/(1-a^2{\cal M}/r^3) <0
\eeq
for the second angular velocity, while the last inequality follows from Eq.~(\ref{eq:sagnacobservers}) and Table 1. Note that both of these angular velocities have the same leading behavior at large $r$, namely $- a{\cal M}/r^3$.
Figure 1 shows the corresponding world lines on the 
$t$-$\phi$ cylinder.

\section{Preliminary considerations}

The Sagnac effect \cite{pos,ash,hen,andbilste,ste1,ste2} and the observer-dependent 
single-clock clock effect (its timelike analog), both
of which in turn are connected to the desynchronization effect,
refer to the asymmetry in the arrival times of a pair of
oppositely rotating null circular orbits 
or timelike geodesic circular orbits at a
given radius as seen by a given rotating observer. These are illustrated in Figure 1 (the Kerr equatorial plane) for the slicing (null meeting point) observers and threading observers (of the original coordinate grid), the geodesic meeting point observers, the geodesic observers, the extremely accelerated observers and the Carter observers. The forward and backward null orbits are labeled by $P_\pm$, while the forward and backward geodesic orbits are labeled by $U_\pm$, all starting at the same spacetime event at the origin of coordinates.

If $(\zeta_1,\zeta_2)$, $\zeta_1 <0< \zeta_2$, is the ordered pair of coordinate angular
velocities of such a pair 
(either $(\zeta_-,\zeta_+)$ for the Sagnac effect or
$(\dot\phi_-,\dot\phi_+)$ for the single-clock clock effect), 
and $\zeta$ is the angular velocity of a
rotating observer with 4-velocity $U$ distinct from this pair, 
and if the observer has an intermediate angular velocity ($\zeta_1<\zeta<\zeta_2$, 
only a restriction in the second case),
one 
finds that
the difference of the coordinate arrival times after one
complete revolution with respect to this observer is 
\begin{eqnarray}
\label{eq:genfor}
   \Delta t 
     &=& {\cal S}(\zeta;\zeta_1,\zeta_2)
       = t_2 - t_1
       =  2\pi\left[ 1/(\zeta_2-\zeta) - 1/(\zeta-\zeta_1) \right]
                \nonumber\\
     &=& -4\pi [ \zeta - (\zeta_1+\zeta_2)/2]
                           / [(\zeta-\zeta_1)(\zeta-\zeta_2)] \ ,
                \nonumber\\
     &=&-2\pi \partial_\zeta \ln [(\zeta-\zeta_1)(\zeta-\zeta_2)] \ .
\end{eqnarray}
If desired the coordinate time difference $\Delta t$ between the two hypersurfaces $t=t_1$ and $t=t_2$ at which the two crossings occur with the observer world line can be converted into a proper time difference $\Delta \tau(U)$ measured by that observer using Eq.~(\ref{eq:lapseparticle}).

To understand the origin of these formulas,
let both orbits and observer start at $t=0$ at the same point in spacetime.
When the corotating orbit first meets the observer at time $t_2$ (equal to the observer-dependent coordinate period of that orbit), the observer has gone a coordinate angle distance $\zeta t_2$ while the corotating orbit has gone a coordinate angle distance $\zeta_2 t_2$, which is one revolution plus the additional amount gone by the observer compared to the coordinate time line through the starting point, i.e. 
$\zeta_2 t_2 = 2\pi + \zeta t_2$ so
$ t_2= 2 \pi/(\zeta_2 - \zeta)$. 
The same consideration for the counterrotating orbit applies except that one must subtract $2 \pi$ from the observer angular distance, obtaining finally
$ t_1= 2 \pi/(\zeta - \zeta_1)$. Their difference is the desired result.

Note that in the limits $\zeta\to\zeta_1$ and $\zeta\to\zeta_2$, the time difference 
${\cal S}(\zeta;\zeta_1,\zeta_2)$ goes to infinity since as the observer angular velocity approaches one of the two orbit values, it takes an increasingly longer time for that faster moving orbit to overtake the slightly slower observer by an extra revolution. 

In terms of relative velocities $\nu(U,n)\equiv \nu $, $\nu(U_{1,2},n)\equiv \nu_{1,2}$ and $\nu_{\rm(mp)}=(\nu_1+\nu_2)/2$ with respect to ZAMOs, the expressions for the general time interval for both the timelike and null cases,
after converting velocity differences with (\ref{eq:nuzetadiffs}),
become
\beq\fl\qquad
\label{eq:time}
{\cal S}(\zeta;\zeta_1,\zeta_2)
 = -4\pi 
\frac{g_{\phi\phi}^{1/2}}{N} 
       \frac{\nu - \nu_{\rm(mp)}}
             {(\nu-\nu_1)(\nu-\nu_2)}\ ,\quad
\label{eq:nul}
{\cal S}(\zeta;\zeta_+,\zeta_-)
 = 4\pi \frac{g_{\phi\phi}^{1/2}}{N} 
\gamma^2 \nu\ .
\eeq
Using these results,
the observer that sees the same time interval both for a pair of oppositely rotating timelike particles and the corresponding photon pair
can be easily found from the condition
\beq
\label{eq:sametime}
{\cal S}(\zeta;\zeta_1,\zeta_2)
={\cal S}(\zeta;\zeta_+,\zeta_-)\ .
\eeq
This reduces to a quadratic equation in the slicing relative velocity of the observer
\beq
\label{eq:oppositespeed}
\nu^2-2\nu \frac{1+\nu_1\nu_2}{\nu_1+\nu_2}+1=0 \ ,
\eeq
which always has two real solutions, one subluminal and the other superluminal \cite{idcf2}.
One can rewrite this equation in the form
\beq\label{eq:addvel}
\frac{\nu_1 -\nu}{1-\nu\nu_1}=-\frac{\nu_2 -\nu}{1-\nu\nu_2} \ ,
\eeq
which is the familiar relativistic addition of velocity formula applied to the two relative velocity differences, equivalent to
\beq
\nu(U_1,U) = -\nu(U_2,U)\ .
\eeq 

Its simple content is that the observer ($U$, corresponding to the subluminal solution $\nu$) that sees the same delay in the arrival times both of the particle and photon pairs also sees the corotating ($U_2$) and counterrotating ($U_1$) particles moving with equal magnitude but oppositely-signed relative velocities. 
But this just means that the observer 4-velocity is the normalized average of the two particle 4-velocities
\beq
\label{eq:zetaUavg}
U=\frac{U_1+U_2}{||U_1+U_2||} \leftrightarrow
\zeta = (\Gamma_1 \zeta_1 + \Gamma_2 \zeta_2)/(\Gamma_1+\Gamma_2)\ .
\eeq
as discussed after Eq.~(4.8) of \cite{idcf2}.

Turning this around,
from the point of view of any fixed observer ($U$), one therefore has a universal time delay for every pair of oppositely rotating timelike circular orbits which are seen symmetrically by that observer (relative velocities of equal magnitude but opposite sign), including the limiting case of the photon pair, independent of the common magnitude of the relative velocities of the pair.

\section{Single-clock clock effect}

For a pair of oppositely rotating timelike geodesics, one has 
\begin{eqnarray}
   \Delta t_{\rm(geo)}(U) 
       & = {\cal S}(\zeta;\dot\phi_-,\dot\phi_+)
         = -4\pi [\zeta - \zeta_{\rm(gmp)}] /
                    [(\zeta-\dot\phi_-)(\zeta-\dot\phi_+)] \nonumber \\
       & = -2\pi \partial_\zeta \ln [(\zeta-\dot\phi_-)(\zeta-\dot\phi_+)]
\ ,
\end{eqnarray}
while for the pair of oppositely rotating null orbits one has
\begin{eqnarray}
   \Delta t_{\rm(null)}(U) 
           & = {\cal S}(\zeta;\zeta_-,\zeta_+)
             = -4\pi [\zeta - \zeta_{\rm(nmp)}] /
                               [(\zeta-\zeta_-)(\zeta-\zeta_+)] 
   \nonumber \\
           &= -2\pi \partial_\zeta \ln [(\zeta-\zeta_-)(\zeta-\zeta_+)]
\ ,
\end{eqnarray}
recalling that 
\begin{equation}
\zeta_{\rm(gmp)} = (\dot\phi_+ +\dot\phi_-)/2 ,\qquad
\zeta_{\rm(nmp)} = (\zeta_+ + \zeta_-)/2 = \zeta_{\rm(sl)} \ .
\end{equation}
Each of these may
be evaluated for any observer $U$. 
Figure 1 illustrates these two time delays for each of the various geometrically defined observers on the equatorial plane of the Kerr spacetime.
Note that
the null arrival time difference is proportional to the angular
momentum (\ref{eq:angmom}) of $U$. 

The difference between the arrival times of circularly rotating timelike geodesics and photons, 
as measured by a generic observer $U$, is simply related to the 
$\zeta$-derivative of the magnitude of the observer 4-acceleration
$A(U)$
\begin{equation}
   \Delta t_{\rm(geo)}(U)-\Delta t_{\rm(null)}(U)=-2\pi \partial_\zeta \ln ||A(U)|| \ ,
\label{eq:difdt}
\end{equation}
since by Eq.~(4.3) of Bini {\it et al}.~\cite{idcf2} one has in the equatorial plane
\begin{equation}
     ||A(U)||
         = |- \kappa(\phi,n)^{\hat r}| 
             \frac{(\zeta-\dot\phi_-)(\zeta-\dot\phi_+)}
                  {(\zeta-\zeta_-)(\zeta-\zeta_+)} \ ,
\end{equation}
where $\kappa(\phi,n)$ is a curvature factor independent of $\zeta$.

This relation gives the extremely accelerated observers 
($\partial_\zeta \ln ||A(U)||=0$) the special property
\beq
\Delta t_{\rm(geo)}(U_{\rm (ext)})=\Delta t_{\rm(null)}(U_{\rm (ext)})\ ,
\eeq
which must be added to their many other special properties arising in the discussion of inertial forces and gyroscope precession measurements \cite{sem0,def}.
In particular the discussion after Eq.~(\ref{eq:sametime}) shows that 
the extremal force observers must therefore see
the oppositely rotating pair of geodesics moving with the same speed.
However, the single-clock clock effect for those observers is negative for Kerr
so the counterrotating geodesic arrives after the corotating one, as
shown in Figure 1.

The Sagnac time difference and the single-clock clock effect associated with the time intervals as seen by the threading observer $U=m$ ($\zeta=0$) are
\begin{eqnarray}
\fl\qquad
     \Delta t_{\rm(null)}(m) 
            &=& {\cal S}(0;\zeta_-,\zeta_+)
             = 4\pi (\zeta_-{}^{-1} + \zeta_+{}^{-1})/2
             = 4\pi M_\phi \ ,
           \nonumber\\
\fl\qquad
     \Delta t_{\rm(geo)}(m) 
            &=& {\cal S}(0;\dot\phi_-,\dot\phi_+)
             = 4\pi (\dot\phi_-{}^{-1} + \dot\phi_+{}^{-1})/2
             = 4\pi a \quad \hbox{[Kerr]} \ ,
\end{eqnarray}
where the final equality only holds for the Kerr spacetime.

The Sagnac effect
$\Delta t_{\rm(null)}(m)$ 
is positive (negative) when the threading observers
corotate (counterrotate) with respect to the slicing observers (${\rm
sgn}\, \Delta t_{\rm(null)} = {\rm sgn}\, N^\phi = {\rm sgn}\,
M_\phi$). This is negative for Kerr (see Table 1), where the slicing observers rotate forward around the hole compared to the threading observers which resist the so called dragging of inertial frames, in contrast with the single-clock clock effect 
$\Delta t_{\rm(geo)}(m)$ 
which is instead positive for Kerr.

The synchronization gap is the time difference which occurs during one spatial loop $C$ along the $\phi$-direction with respect to the threading observers. Integrating the differential condition of Eq.~(\ref{eq:dtM}) along a counterclockwise trip (increasing $\phi$) one finds
\begin{equation}
\label{eq:sagn}
 \fl\qquad   
\Delta t_{\rm(SG)}(m) = \int_C dt
   =\int_0^{2\pi} M_\phi\, d\phi
                 = 2\pi M_\phi = 2\pi / \bar\zeta_{\rm (th)} 
           =\Delta t_{\rm(null)}(m) /2 \ .
\end{equation}
The synchronization gap is exactly half the Sagnac time difference  (see Figure 1), the factor of 2 coming from the fact that the Sagnac loop consists of 2 revolutions compared to the single revolution of the synchronization gap `loop.' 
To understand this close connection, one must also approach the Sagnac time difference by integrating a differential condition, as will be done below. Note that this relation is true for a generic observer taken as the new threading observer, and so holds for the corresponding tilde quantities
\begin{equation}
\label{eq:Usagn}
 \fl\qquad   
\Delta \tilde t_{\rm(SG)}(U) = \int_{\tilde C} d\tilde t
   =\int_0^{2\pi} \tilde M_{\tilde\phi}\, d\tilde\phi
                 = 2\pi \tilde M_{\tilde\phi}
           =\Delta \tilde t_{\rm(null)}(U) /2 \ .
\end{equation}

The analogous coordinate time difference for the geodesic case in the Kerr spacetime is the single-clock clock effect for the threading observers
\begin{equation}
\label{eq:clock}
   2\pi / \bar\zeta_{\rm(car)} = \Delta t_{\rm(geo)}(m) /2
       = 2\pi a
\end{equation}
and corresponds exactly to one threading loop of a circular curve which is
spatial instead with respect to a Carter observer
connecting the average-time-of-return point on the threading observer
world line with either return point on the same world line.
Thus the Carter observers in the single-clock clock effect correspond to the threading observers in the Sagnac effect in a certain sense. 
Figure 1 shows this loop, described by $t' = t-a\phi=0$, where $t'$ is a local helical time coordinate orthogonal to these observers in the equatorial plane only \cite{greschvis}. With $\phi' =\phi$, one has a new local slicing for which $d\phi'/dt' = \zeta/(1-a\zeta)=y$ is the angular velocity variable of de Felice
\cite{def,def0}.

One can express the threading observer synchronization gap, Sagnac
effect, and single-clock clock effect with respect to the threading observers using a single
formula in terms of either the angular velocity or threading relative velocity. The 1-form 
\begin{eqnarray}
\fl\qquad
  [\nu(U,m)\gamma(U,m)]^{-1} \bar U_\alpha dx^\alpha
   &=& - M (dt - \zeta^{-1} d\phi)           \nonumber\\
\fl\qquad
    &=& - M (dt - M_\phi d\phi)  
        + \gamma_{\phi\phi}{}^{1/2}/\nu(U,m)^{\hat\phi} \, d\phi
\end{eqnarray}
orthogonal to the 4-velocity $U$ restricts to zero its
world line segment $C$, so the threading observer proper travel time (proper period as seen by the observer) is 
\begin{eqnarray}
\fl\qquad      \tau(m) &=& M  t(m)
           = M \int_C dt = [{\rm sgn}\,\zeta] M \int_0^{2\pi} \zeta^{-1} d\phi   = 2\pi M/|\zeta| 
   \nonumber\\ 
           &=& [{\rm sgn}\, \nu(U,m)^{\hat\phi}]
              \int_0^{2\pi} [MM_\phi 
        + \gamma_{\phi\phi}{}^{1/2}/\nu(U,m)^{\hat\phi}] \, d\phi
             \nonumber\\
\fl\qquad       &=&
               2\pi [{\rm sgn}\, \nu(U,m)^{\hat\phi}] [ MM_\phi 
        + \gamma_{\phi\phi}{}^{1/2}/\nu(U,m)^{\hat\phi}]
\ .
 \end{eqnarray}
Thus the difference in arrival times of two oppositely rotating orbits is then 
\begin{eqnarray}
\fl\qquad
    \Delta \tau(m) &=& M[t_2(m) - t_1(m)] \nonumber\\
\fl\qquad
     &=& 2\pi \left[2MM_\phi + \gamma_{\phi\phi}{}^{1/2}
               [1/\nu(U_1,m)^{\hat\phi} + 1/\nu(U_2,m)^{\hat\phi}]
            \right]
         \ ,
\end{eqnarray} 
correcting Eq.~(7.10) of \cite{idcf2} by a misplaced factor of 2.

These same considerations apply to a generic stationary
observer ($U$) and any circular orbit ($U_1$), once they are expressed in terms of the new threading potentials and fields
\beq\label{eq:genclock1}
\fl      
\tau_1(U) = \tilde M  \tilde t_1(U)
           = \tilde M \int_{\tilde C} d\tilde t 
 = [{\rm sgn}\,(\zeta_1-\zeta)] 
  \tilde M \int_0^{2\pi} 1/(\zeta_1-\zeta)\, d\tilde \phi  
 =  2\pi \tilde M/|\zeta_1-\zeta|  \ .
\eeq
As in the original threading case, the difference in arrival times of two oppositely rotating orbits ($U_1$ and $U_2$) can be expressed in terms of the observer relative velocities as
\begin{eqnarray}\label{eq:genclock2}
\fl\qquad
    \Delta \tau(U) &=& \tilde M[\tilde t_2(U) - \tilde t_1(U)]
 \nonumber\\
\fl\qquad
     &=& 2\pi \left[2\tilde M \tilde M_{\tilde\phi} + 
\tilde\gamma_{\tilde\phi\tilde\phi}{}^{1/2}
               [1/\nu(U_1,U)^{\hat{\tilde\phi}} + 1/\nu(U_2,U)^{\hat{\tilde\phi}}]
            \right]
         \ .
\end{eqnarray} 
For relative velocities of equal magnitude but opposite sign, this reduces to the first term alone, which is the proper time synchronization gap $2\Delta \tau_{\rm(SG)}(U)$. 
For photons this is always true for any observer, identifying the Sagnac effect with the synchronization gap (for two loops) for that observer.
Thus for the constant angular velocity orbits, the one loop effect is just the local effect in the tangent space illustrated in Figure 2, multiplied by the total angle $2\pi$ of the loop.

Similarly,  
the extremely accelerated observers see the oppositely rotating circular geodesics with equal magnitude but
oppositely-signed relative velocities, so this difference in the arrival times that they measure is also entirely due to their own synchronization gap. 
This is in contrast with the Carter observers whose synchronization gap over a complete revolution with respect to the threading observers equals the single-clock clock effect for those latter observers, and so is a hybrid quantity.

\section{Two-clock clock effects}

The two-clock clock effects, towards which recent attention has been directed \cite{masgrolic,bonste,sem,tart1,tart2},
can now be calculated easily from the formula (\ref{eq:genclock1}), corrected for the new proper times. The period of the orbit of $U_1$ in terms of the proper time of $U_1$ only differs by replacing $\tilde M$ by $\Gamma_1^{-1}$ in Eq.~(\ref{eq:genclock1}).
The two proper periods of the oppositely rotating geodesics ($U_\pm$) after one revolution with respect to $U$
are therefore
\beq\label{eq:genclock1b}
 \tau(U_\pm,U) =  \tilde t_\pm(U)/\Gamma_\pm
  =  2\pi \Gamma_\pm^{-1}/|\dot\phi_\pm-\zeta|  \ .
\eeq
The difference in the two proper periods of the oppositely rotating geodesics ($U_\pm$) then becomes
\begin{eqnarray}\label{eq:genclock2b}
\fl
    \Delta \tau(U_+,U_-,U) &=& \tilde t_+(U)/\Gamma_+ - \tilde t_-(U)/\Gamma_- 
= 2\pi [ \Gamma_+^{-1}/ (\dot\phi_+ -\zeta) 
            - \Gamma_-^{-1}/ (\zeta-\dot\phi_-) ]\nonumber\\
 \fl
 &=& 2\pi g_{\phi\phi}^{1/2} (\gamma_+^{-1} + \gamma_-^{-1})
       \frac{\nu(U,n)-\nu(U_{\rm(ext)},n)}
             {[\nu(U,n)-\nu(U_+,n)][\nu(U,n)-\nu(U_-,n)]}
         \ ,
\end{eqnarray} 
where in the last equation the angular velocities have been expressed in terms of the slicing relative velocities using (\ref{eq:zetanuUmn})
and
Eq.~(\ref{eq:zetaUavg}) for the angular velocity of the extremely accelerated observers has been used \cite{idcf2} . Thus although they have different arrival times, the two oppositely rotating geodesics measure the same proper time period after one revolution with respect to the extremely accelerated observers, as shown by Semer\'ak for the Kerr and van Stockum spacetimes \cite{sem}. 
This is true for any stationary axisymmetric spacetime, including the G\"odel spacetime, which also has extremely accelerated observers \cite{idcf2}. 

Finally the observer-independent two-clock clock effect results from specializing these equations to the geodesic meeting point observer, recovering the results of Tartaglia \cite{tart1,tart2} for the Kerr and van Stockum 
spacetimes. 
The proper periods between three successive crossings corresponding to each geodesic making a complete revolution 
(with respect to such an observer)
are
\beq\label{eq:genclock1c}
\fl\qquad
 \tau(U_\pm,U_{\rm(gmp)}) =  \tilde t_\pm(U_{\rm(gmp)})/\Gamma_\pm
  =  2\pi \Gamma_\pm^{-1}/|\dot\phi_\pm-\zeta_{\rm(gmp)}| 
  =  4\pi \Gamma_\pm^{-1}/|\dot\phi_+ -\dot\phi_-| 
\eeq
and their difference is
\beq\label{eq:genclock2c}
 \Delta\tau(U_+,U_-,U_{\rm(gmp)})
  = 4\pi [\Gamma_+^{-1} -\Gamma_-^{-1}]/[\dot\phi_+ - \dot\phi_-] \ .
\eeq
The latter result must be halved to compare with the result of Tartaglia \cite{tart1,tart2} for Kerr, since he considers two successive crossings of the geodesics.

Notice also that the first meeting point after the pair of geodesics leave $\phi=0$ will be at $\pi$ minus half the amount of angle through which the geodesic meeting point observers have moved in one full revolution
\beq
  \phi_{\rm(gmp)} = \pi +\frac12 \zeta_{\rm(gmp)} \tilde t_\pm(U_{\rm(gmp)})
         = \pi\left[ 1 
    + \frac{\dot\phi_+ + \dot\phi_-}{\dot\phi_+ - \dot\phi_-} \right] \ .
\eeq 
If $\dot\phi_+ < |\dot\phi_-|$, as occurs in the Kerr case, the geodesic meeting point observers counterrotate with respect to the threading observers, since the counterrotating geodesics are moving faster with respect to the threading observers. This goes against the naive `dragging of inertial frames' intuition \cite{rin,masgrolic,masgrothe}.
The total angular difference after $n$ full revolutions would then be
\beq
  \Delta\phi_{\rm(gmp)} = n \zeta_{\rm(gmp)} \tilde t_\pm(U_{\rm(gmp)})
         =  2 n\pi
    \frac{\dot\phi_+ + \dot\phi_-}{\dot\phi_+ - \dot\phi_-}\ ,
\eeq 
which is another quantity one might consider measuring.



\begin{figure}
\typeout{*** EPS figure 3}
\centerline{
\epsfbox{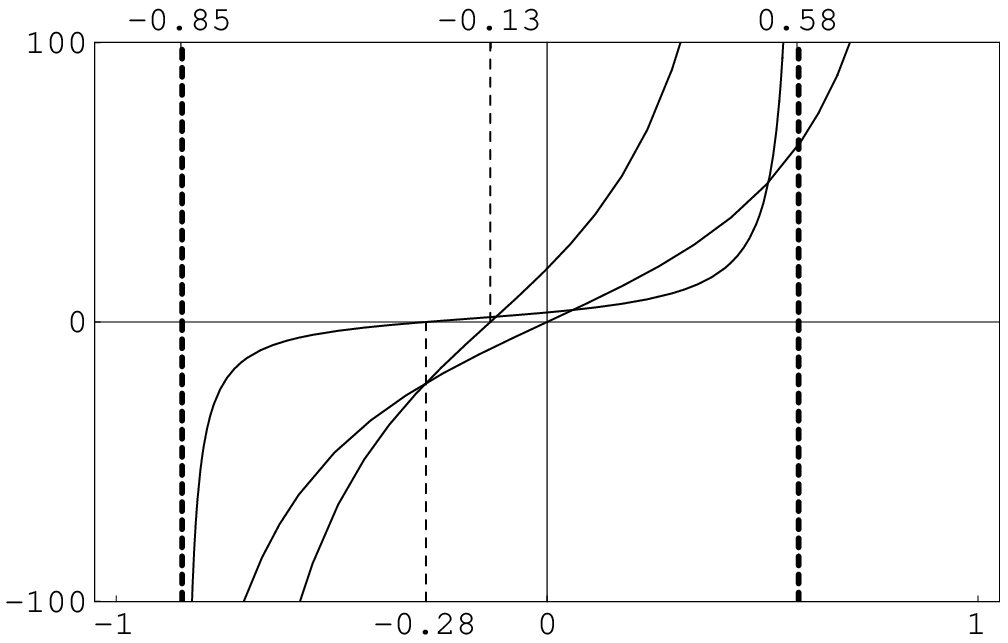}
}
\vspace{1.5cm}
\caption{The behaviour of 
$\Delta \tau(U_+,U_-,U)$ (top right, middle curve), 
$\Delta t_{\rm (geo)} (U)$ (top right, left curve),
and $\Delta t_{\rm (null)} (U)$ (top right, right curve),
at $r=4$ is shown as a function of the slicing relative velocity $\nu(U,n)$
in the case of circular orbits in the equatorial plane of the Kerr spacetime with  $a=0.5$, ${\cal M}=1$.
The two vertical asymptotes of $\Delta t_{\rm (geo)} (U)$ correspond to the corotating and counterrotating geodesic velocities.
Velocity values are indicated which correspond to zero values of  
$\Delta \tau(U_+,U_-,U)$ (extremely accelerated observers: 
$\nu(U,n)=-0.28$),
$\Delta t_{\rm (geo)} (U)$ (geodesic meeting point observers:
$\nu(U,n)=-0.13$),
and 
$\Delta t_{\rm (null)} (U)$ (ZAMOs: 
$\nu(U,n)=0$). 
The crossing condition  
$\Delta t_{\rm (geo)} (U)=\Delta t_{\rm (null)} (U)$ 
also occurs at the extremely accelerated observer velocity.
}
\label{fig:2}
\end{figure}


Figure 3 compares the three clock effects for a typical radius in the equatorial plane of Kerr spacetime.

\section{Explicit examples}

\subsection{G\"odel}

For the G\"odel spacetime the $t$-$\phi$ 2-metric is
\beq
       ds_{(t,\phi)}^2 
       = -dt^2 + 4\Omega^{-1} s^2 \, dt d\phi 
          + 2 \Omega^{-2} s^2(1-s^2)\, d\phi^2\ ,
\eeq 
where $s=\sinh(\Omega r/\sqrt{2})$ and $\Omega$ is the vorticity parameter of its fluid source \cite{idcf2}. 
The geodesic angular velocities and coordinate gamma factors are
\begin{eqnarray}
  \dot\phi_+ &=& 0\ ,\ \dot\phi_- = -2\Omega/(1-2s^2)\ ,\nonumber\\ 
  \Gamma_+ &=& 1\ ,\ \Gamma_- = \frac{1-2s^2}{\sqrt{1-4 s^2 (1+s^2)}}\ ,
\end{eqnarray}
while the angular velocity of the extremely accelerated observer is
\beq
  \zeta_{\rm(ext)} 
     = -\frac{\Omega}{4 s^4} [ 1-2s^2 - \sqrt{1-4 s^2 (1+s^2)}]\ .
\eeq
With these formulas, both the observer-dependent two-clock clock effect 
and the observer-independent two-clock clock effect then follow from Eqs.~(\ref{eq:genclock2b}) and (\ref{eq:genclock2c}).

\subsection{Kerr-Taub-NUT equatorial plane}

For the Kerr-Taub-NUT spacetime in the equatorial plane, the $t$-$\phi$ 2-metric coefficients are \cite{mil}
\beq
\fl\qquad
g_{tt} = -1+2 W \ ,\
g_{t\phi} = -2a W\ ,\
g_{\phi\phi} = (r^2+a^2+\ell^2) +2 a^2 W \ ,
\eeq
where
$W = ({\cal M}r+\ell^2)/(r^2+\ell^2)$
and $\ell$ is the Taub-NUT parameter, which may be interpreted as the gravitomagnetic monopole moment of the source of the gravitational field or in other ways \cite{bon,bra} and $J={\cal M}a$ is the angular momentum, namely the gravitomagnetic dipole moment of the source.

The geodesic angular velocities are
\begin{eqnarray}
\fl\qquad
  \dot\phi_\pm^{-1} &=& a(1+\frac{2\ell^2r}{{\cal M}(r^2-\ell^2)})        \nonumber\\ 
\fl\qquad
& &
          \pm \sqrt{\frac{r^3}{\cal M}}
  \left[\frac{(r^2+\ell^2)^2}{r^2(r^2-\ell^2)} + \frac{2a^2\ell^2}
         {r^2(r^2-\ell^2)} \left(1+ \frac{2\ell^2r}{{\cal M}(r^2-\ell^2)} \right)\right]^{1/2}
\ ,
\end{eqnarray}
which in the limit $r\to\infty$, together with their
coordinate gamma factors, become 
\begin{eqnarray}
  \dot\phi_\pm^{-1} &\approx& a \pm \sqrt{\frac{r^3}{\cal M}} 
         \pm \frac32 \frac{\ell^2}{\sqrt{{\cal M}r}}
           + \frac{2a\ell^2}{{\cal M}r} \ ,
        \nonumber\\ 
\Gamma_\pm &\approx& 1+ \frac32\frac{\cal M}{r} + \frac{\ell^2}{r^2}
   + \frac{27}{8} \frac{{\cal M}^2}{r^2} 
  \mp 3 \frac{a}{r} \left( \frac{\cal M}{r} \right)^{3/2}
\ ,
\end{eqnarray}
while the null meeting point,  geodesic meeting point and extremely accelerated observer angular velocities are
\begin{eqnarray}
  \zeta_{\rm(nmp)} 
     &\approx& - \frac{a{\cal M}}{r^3} 
   \left[ 2  \phantom{ + \frac{3{\cal M}}{r}\ \ } +\frac{2\ell^2}{{\cal M}r} \right] \ ,
  \nonumber\\
  \zeta_{\rm(gmp)} 
     &\approx& - \frac{a{\cal M}}{r^3} 
   \left[ 1 \phantom{ + \frac{3{\cal M}}{r}\ \ } +\frac{2\ell^2}{{\cal M}r} \right] \ ,
  \nonumber\\
  \zeta_{\rm(ext)} 
     &\approx& - \frac{a{\cal M}}{r^3} 
   \left[ 1 + \frac{3{\cal M}}{r} +\frac{2\ell^2}{{\cal M}r} \right]
\ .
\end{eqnarray}

With these formulas, the various clock effects may be evaluated in this limit
\begin{eqnarray}
    \Delta \tau_{\rm(geo)}(m)      
&\approx&  \phantom{-\,\,}4\pi a 
   \left[ 1  \phantom{+ \frac{3{\cal M}}{r} +\frac{6{\cal M}^2}{r^2}\ \ }   
  + \frac{2\ell^2}{{\cal M}r} \right]\ ,
  \nonumber\\
 \Delta\tau_{\rm(geo)}(U_{\rm(ext)})
 &\approx& -4\pi a
 \left[ 0 + \frac{3{\cal M}}{r} +\frac{6{\cal M}^2}{r^2} + \frac{4\ell^2}{r^2} \right]
  \ ,\nonumber\\
    \Delta \tau(U_+,U_-,m)      
&\approx&  \phantom{-\,\,}4\pi a 
   \left[ 1+ \frac{3{\cal M}}{2r} \phantom{+\frac{6{\cal M}^2}{r^2}\ \ } +\frac{2\ell^2}{{\cal M}r} \right] \ ,
  \nonumber\\
 \Delta\tau(U_+,U_-,U_{\rm(gmp)})
 &\approx& \phantom{-\,\,}4\pi a
 \left[ 0 + \frac{3{\cal M}}{r} +\frac{9{\cal M}^2}{2r^2} + \frac{4\ell^2}{r^2} \right] \ .
\end{eqnarray} 
Notice that the difference between the single-clock and two-clock threading clock effects as well as the extremely accelerated observer 
single-clock effect 
and the clock-independent effect 
are suppressed by the factor ${\cal M}/r$, so experimentally there is really only one clock effect to be measured for solar system scenarios. Moreover,
it turns out that the angular difference 
\beq
\Delta\phi_{\rm(gmp)}
  \approx -\pi a \sqrt{\frac{{\cal M}}{r^3}} 
   \left[ 1 +\frac{2\ell^2}{{\cal M}r} -\frac{3\ell^2}{2r^2}\right]
\eeq
is negligibly small in realistic experimental situations in the solar system.

The fact that both the gravitomagnetic
monopole and dipole moments of the source of the gravitational field for this spacetime are nonzero makes it an interesting example for examining the clock effects. Notice that only $\ell^2$ enters the metric in the equatorial plane, so the sign of $\ell$ is unimportant for these effects. For $r>>{\cal M}$ and $r>>\ell^2/{\cal M}$ the threading observer clock effects behave like 
\beq
4\pi a = 4\pi \frac{J}{{\cal M}c^2} \ , 
\eeq
which is independent of the gravitational constant $G$ and the orbital radius $r$ 
\cite{cohmas,masgrothe,bonste,massan}
and has a value of about $10^{-7}$ sec for the Earth. However, there are experimental difficulties that must be overcome before there is any hope of measuring this effect \cite{licgromas,tart1,tart2,ior}.

Note that the gravitomagnetic clock effect may be extended to general orbits based on the idea of `azimuthal closure' \cite{cohmas,masgrothe,bonste,massan}. This also makes it clear that the definition of the azimuthal angle is crucial for the determination of this effect.  In principle one can use it to measure the rotation rate 
of the astronomical reference frame that is used for the interpretation of observations relative to the underlying spacetime with respect to which the source has intrinsic specific angular momentum $a$ \cite{wol}.

\section{Conclusion}

The various gravitomagnetic clock effects, the Sagnac effect and the desynchronization effect for circularly rotating orbits in stationary axisymmetric spacetimes are all closely related.
A relative observer analysis has shown how these effects can be studied together and how they are encoded into the symmetry adapted coordinates.  
Special observer families then naturally arise. 
Extremely accelerated observers are shown to be the observer family for which the single-clock clock effect and Sagnac effects both agree since they both reduce to the synchronization gap, 
while their two-clock clock effect vanishes as previously shown by Semer\'ak. On the other hand the geodesic meeting point observers are those for which the single-clock clock effect vanishes in direct analogy with the locally nonrotating observers (null meeting point observers) for which the Sagnac effect vanishes. 

\section*{Acknowledgments}

The authors are grateful to the organizers of the Spanish Relativity Meeting EREs2000 that
led to this collaboration.

\end{document}